# Telescope interferometers: an alternative to classical wavefront sensors

F. Hénault

UMR 6525 CNRS H. FIZEAU – UNS, OCA, Avenue Nicolas Copernic, 06130 Grasse – France

## ABSTRACT

Several types of Wavefront Sensors (WFS) are nowadays available in the field of Adaptive Optics (AO). Generally speaking, their basic principle consists in measuring slopes or curvatures of Wavefront Errors (WFE) transmitted by a telescope, subsequently reconstructing WFEs digitally. Such process, however, does not seem to be well suited for evaluating co-phasing or piston errors of future large segmented telescopes in quasi real-time. This communication presents an original, recently proposed technique for direct WFE sensing. The principle of the device, which is named "Telescope-Interferometer" (TI), is based on the addition of a reference optical arm into the telescope pupil plane. Then incident WFEs are deduced from Point Spread Function (PSF) measurements at the telescope focal plane. Herein are described two different types of TIs, and their performance are discussed in terms of intrinsic measurement accuracy and spatial resolution. Various error sources are studied by means of numerical simulations, among which photon noise sounds the most critical. Those computations finally help to define the application range of the TI method in an AO regime, including main and auxiliary telescope diameters and magnitude of the guide star. Some practical examples of optical configurations are also described and commented.

**Keywords:** Wave-front sensing, Fourier optics, Telescope-Interferometers, Phase measurement, Astronomical optics

## 1. INTRODUCTION

The principle of Adaptive Optics (AO) was first proposed by Babcock [1] in 1953, and encountered continually growing success after a few decades. Today the largest observatories on Earth are all equipped with this technology that demonstrates outstanding capacities to pass beyond the seeing limit and reveal unsuspected details about numerous types of sky objects, particularly in the infrared region of the electromagnetic spectrum. But adaptive optics is continually faced to new challenges, such as covering low-wavelength spectral domain, or attaining extreme Strehl ratios for the detection of extra-solar planets with new generation, planet-finding instruments [2], where more than one thousand mirror actuators are needed. In addition, AO has to cope with the increasing size of future Extremely Large Telescopes (ELTs), with diameters ranging from 30 to 50 meters. For such facilities, it is expected that the primary mirror will be made of an array of smaller reflecting segments, like the Keck, GranTeCan or JWST (James Webb Space Telescope) already are. In that case, one of the most critical problems becomes to adjust (or co-phase) the individual pistons of the segments in order to approximate the continuous theoretical surface of the primary mirror within accuracies typically better than one tenth of wavelength. Looking deeper into the future, the imaging hyper-telescope proposed by Labeyrie [3] will also impose to develop robust co-phasing capacities.

The measurement of the Wavefront Error (WFE) emerging from a telescope can be carried out in several different ways. In the field of AO, the most common method is to sense WFE by means of a pupil plane Wavefront Sensor (WFS), such as Shack-Hartmann [4], curvature [5], pyramidal [6] or optical differentiation sensors [7]. In the most general case however, these devices are not suitable for co-phasing mirror segments, because their basic principle consists in measuring phase slopes and then retrieving WFE using digital procedures. Hence they do not recognize piston errors. To overcome this difficulty, another way is to employ image plane restoration techniques such as phase retrieval [8] or phase diversity [9]. However, such processes are not well matched to AO operation because they usually require significant post-processing times. The "ideal" wavefront sensor should indeed combine advantages of both methods, i.e. the ability to perform direct WFE measurements in quasi real-time. Such a new generation of WFS is already rising however. Let us mention as examples the works from Angel [10] and Labeyrie [11], whose ideas are to move the WFS from the pupil down to the image plane, where its design would be based on a Mach-Zehnder interferometer, eventually using holographic techniques.

Recently, a new approach was suggested for direct WFE sensing [12-13], combining some of the previous trends since the measurement is directly achieved at the telescope focal plane, on the one hand, and the telescope is equipped with an additional, specific module, on the other hand. Indeed, the idea consists in transforming the telescope itself into a phase sensing apparatus by creating sets of interference fringes into the focal plane. Practically, this is realized by adding one reference arm at the pupil plane of the telescope – a modification requiring subsequent additional opto-mechanical hardware. Information about WFE is then extracted from the measured signal by means of reasonably simple data processing algorithms. The general principles of such a device, which I called "Telescope-Interferometer" (TI) are summarized in section 2, where two different TI families are described. Then a general error analysis of both types of TIs is provided in section 3, including random noise as well as systematic errors (or bias). Typical applications and their range of validity are also discussed in that section, and a preliminary trade-off between both types of TIs is conducted. Finally, two examples of practical implementation of phase-shifting TIs are described in section 5.

## 2. THEORY

Two different types of Telescope-Interferometers were described so far, namely the off-axis and phase-shifting TIs. Their basic theories are detailed in Refs. [12] and [13] respectively, and the reader is invited to look at those papers. In order to present a self-content communication however, I provide herein two short and alternative approaches, based on the direct evaluation of Optical Transfer Functions (OTFs) by means of cross-correlations. In any case, the fundamental principle of the TI method consists to:

- acquire one or several Point-Spread Functions (PSFs) at the telescope focal plane, then
- compute numerically their associated OTFs by means of an inverse Fourier transform, and finally
- retrieve the Wavefront Error of the telescope from the phase(s) of the OTF(s).

That procedure is suitable to Adaptive Optics applications, since the used algorithms are rather simple and compatible with quasi real-time operation. The major difference between the off-axis and phase-shifting TI resides in types and shapes of the TI output pupils, such as depicted in Fig. 1. In both cases we consider a main telescope aperture of diameter $D = 2R$ and full geometrical area $S_R$, and an additional aperture, named "reference pupil", of diameter $d = 2r$ and geometrical area $S_r$. The key hypothesis consists in assuming that r is significantly smaller than R, i.e. the ratio $C = S_r/S_R$ is negligible with respect to unity ($C << 1$). C is one basic characteristic of any telescope-interferometer, called its "contrast ratio". In the next sections are used the following scientific notations: $k = 2\pi/\lambda$ where $\lambda$ is the wavelength of the incoming light (supposed to be monochromatic), and $\otimes$ and $*$ respectively stand for the cross correlation and convolution products.

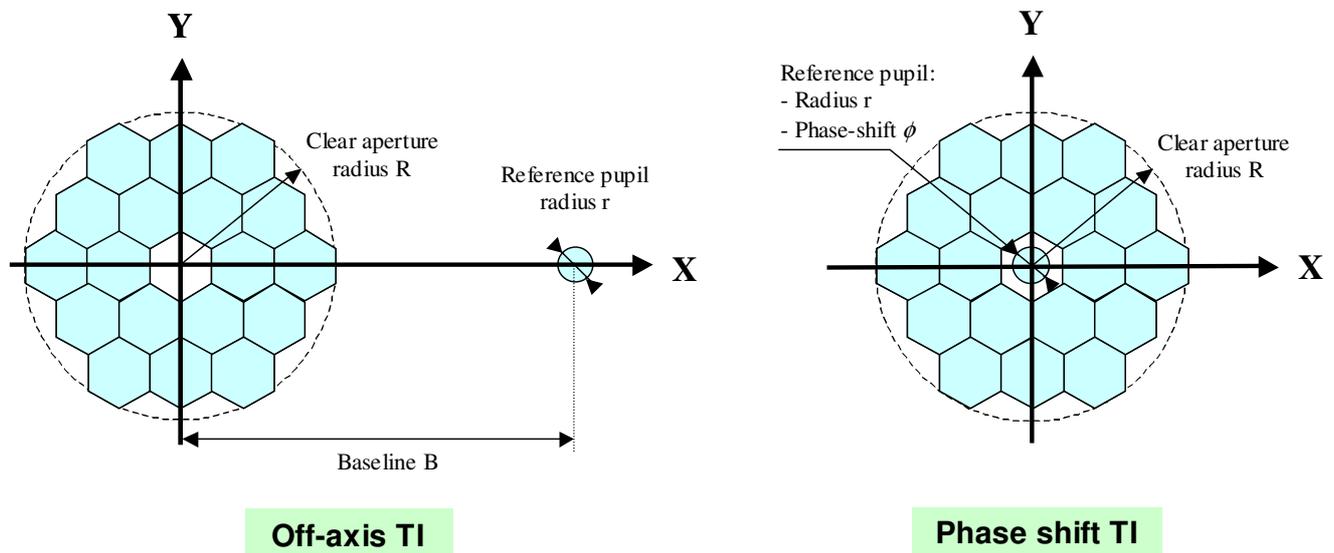

Fig. 1. Useful pupil areas for the off-axis (left) and phase-shifting TIs (right).

## 2.1 Off-axis Telescope-Interferometer

In that configuration, the reference pupil is de-centred of a distance B with respect to the optical axis of the main telescope (see Fig. 1, left side). Let us denote $A(\vec{P})$ the complex amplitude in the pupil plane, as a function of vector $\vec{P}$ of coordinates (x,y). The pupil function of the off-axis TI can be written:

$$A(\vec{P}) = B_R(\vec{P})\exp[ik\Delta(\vec{P})] + B_r(\vec{P} - \vec{P}_0) \quad (1)$$

with $\Delta(\vec{P})$ being the WFE to be measured, and $\vec{P}_0$ the "baseline" vector of coordinate (B,0). $B_R(\vec{P})$ is the two-dimensional transmission function of a circular pupil of radius R, uniformly equal to one inside the rim, and to zero anywhere else – it must be noticed that this function may not be really circular (as in Fig. 1), therefore R stands for the clear aperture of the main telescope. Conversely $B_r(\vec{P})$ is the transmission map of the reference pupil, equal to the "top-hat" function of radius r. We search for a mathematical expression of $C_P(\vec{P})$, which is the inverse Fourier transform of the measured PSF at the TI focal plane. But $C_P(\vec{P})$ can also be considered as the OTF of the off-axis TI, such that [14]:

$$C_P(\vec{P}) = A(\vec{P}) \otimes A(\vec{P}) \quad (2)$$

Here will be invoked some classical properties of cross correlation and convolution products, and in particular:

$$U(\vec{P}) \otimes V(\vec{P}) = U^*(-\vec{P}) * V(\vec{P}) \quad (3)$$

whatever the functions $U(\vec{P})$ and $V(\vec{P})$ – and superscript * stands for complex conjugates. Owing to the fact that both $B_R(\vec{P})$ and $B_r(\vec{P})$ are centro-symmetric, $C_P(\vec{P})$ can be developed as follows:

$$C_P(\vec{P}) = B_R(\vec{P})\exp[ik\Delta(\vec{P})] \otimes B_R(\vec{P})\exp[ik\Delta(\vec{P})] + B_r(\vec{P}) \otimes B_r(\vec{P})$$
$$+ B_R(-\vec{P})\exp[-ik\Delta(-\vec{P})] * B_r(\vec{P} - \vec{P}_0) + B_R(\vec{P})\exp[ik\Delta(\vec{P})] * B_r(\vec{P} + \vec{P}_0) \quad (4)$$

Fig. 2 shows a typical illustration of one measured PSF at the focus of an off-axis TI under favorable atmospheric conditions (Fried's radius $r_0$ is equal to 50 mm), while Fig. 3 represents the modulus of the derived OTF, which is composed of four different terms as predicted by Eq. (4).

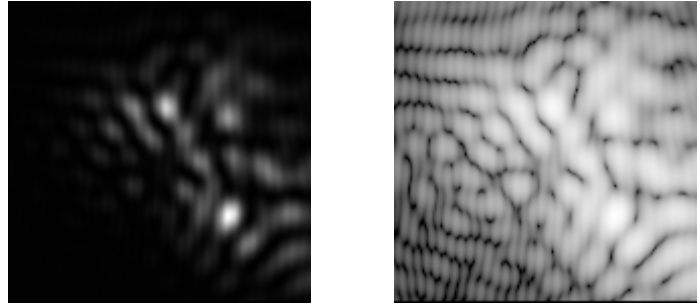

Fig. 2. Example of telescope PSF generated by an off-axis TI in the presence of atmospheric perturbations with Fried's radius $r_0$ = 50 mm (left: linear scale; right: logarithmic scale). A weak fringes modulation is clearly visible.

The first two terms of Eq. (4) are proportional to the OTFs of the main and reference pupils. Denoting them $OTF_R(\vec{P})$ and $OTF_r(\vec{P})$ respectively and dividing Eq. (4) by $S_R$, we get:

$$C_P(\vec{P}) = OTF_R(\vec{P}) + C \times OTF_r(\vec{P}) + C \times B_R(-\vec{P})\exp[-ik\Delta(-\vec{P})] * B_r(\vec{P} - \vec{P}_0)/S_r$$
$$+ C \times B_R(\vec{P})\exp[ik\Delta(\vec{P})] * B_r(\vec{P} + \vec{P}_0)/S_r \quad (5)$$

The fourth term is easily isolated from the other and re-centred on the origin:

$$C_P(\vec{P} - \vec{P}_0) \times B_{R+r}(\vec{P}) = C \times B_R(\vec{P})\exp[ik\Delta(\vec{P})] * B_r(\vec{P})/S_r \quad (6)$$

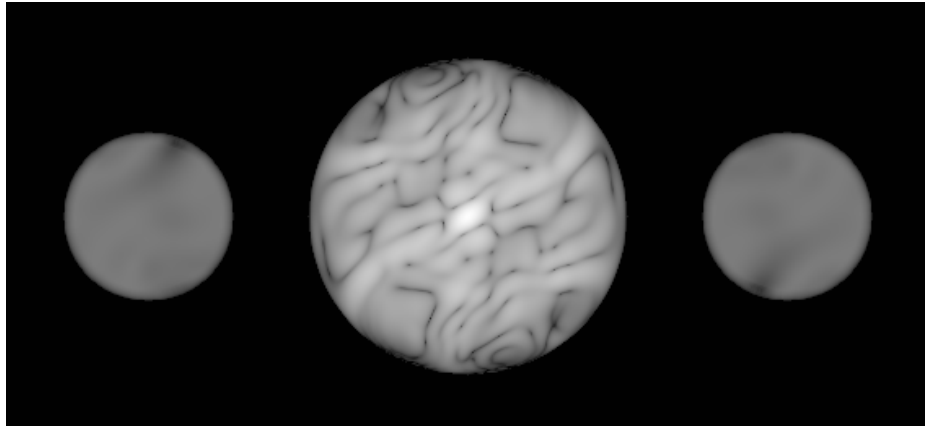

Fig. 3. Example of Modulation Transfer Function (MTF) produced by an off-axis TI in the presence of seeing with Fried's radius $r_0$ = 50 mm (logarithmic scale). Two symmetric "satellites" images of the main pupil appear, whose phase is proportional to the telescope WFE. Obviously the method remains valid as long as there is no overlaps with the central term, one condition that is always fulfilled if B > 3R + r [12].

Here is introduced the so-called "Delta approximation", which constitutes together the main strength and weakness of the method, as will be discussed further. It consists in assuming that function $B_r(\vec{P})/S_r$ tends towards the Dirac distribution $\delta(\vec{P})$ as $S_r$ gets significantly lower than $S_R$. It implies that:

$$B_R(\vec{P})\exp[ik\Delta(\vec{P})] \approx C_P(\vec{P}-\vec{P}_0) \times B_{R+r}(\vec{P})/C \qquad (7)$$

Eq. (7) is the final phase retrieval formula applicable to an off-axis TI, demonstrating that the searched Wavefront Error $\Delta(\vec{P})$ is proportional to the phase of the complex number $C_P(\vec{P}-\vec{P}_0)$. As a major consequence, there remains a $2\pi$-ambiguity on the phase value, thus the retrieved WFE will be enclosed in a $\pm\lambda/2$ range. That statement is also valid for phase-shifting Telescope-Interferometers. Most generally $2\pi$-ambiguities can be removed by means of classical phase-unwrapping algorithms, however in some specific cases (such as determination of piston errors of a large segmented mirror), some additional measurements performed at different wavelengths might be combined. Fig. 4 and Fig. 5 present two examples of numerical simulations where phase unwrapping is efficient. In both cases $\lambda$ is equal to 0.633 µm, and the main telescope and reference pupil diameters are D = 2R = 5 m and d = 2r = 0.5 m respectively. The aperture number of the main telescope is 10. Here the two original WFEs are showing moderate atmospheric perturbations with $r_0$ = 25 mm (Fig. 4), on the one hand, and high-spatial frequency polishing errors of a large telescope mirror that was actually manufactured (Fig. 5), on the other hand. Attained performance is indicated in both Figures, which clearly illustrate the phase retrieval procedure – showing moduli of crossed OTF term, rough (or wrapped) crossed-term phase, and final reconstructed WFEs as well as their difference maps with respect to original WFEs for comparison purpose.

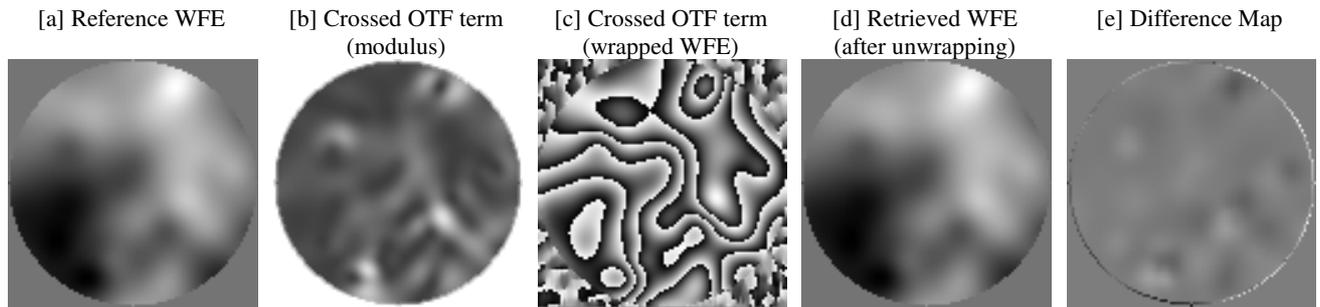

Fig.4. Case of an off-axis TI sensing atmospheric perturbations ($r_0$ = 25 mm). [a] Reference WFE – PTV = 6.811 $\lambda$; RMS = 1.497 $\lambda$ with $\lambda$ = 0.6328 µm. [d] Reconstructed WFE – PTV = 6.781 $\lambda$; RMS = 1.495 $\lambda$. [e] Bi-dimensional difference-map – PTV = 0.159 $\lambda$; RMS = 0.008 $\lambda$. Grey-levels are scaled to PTV values.

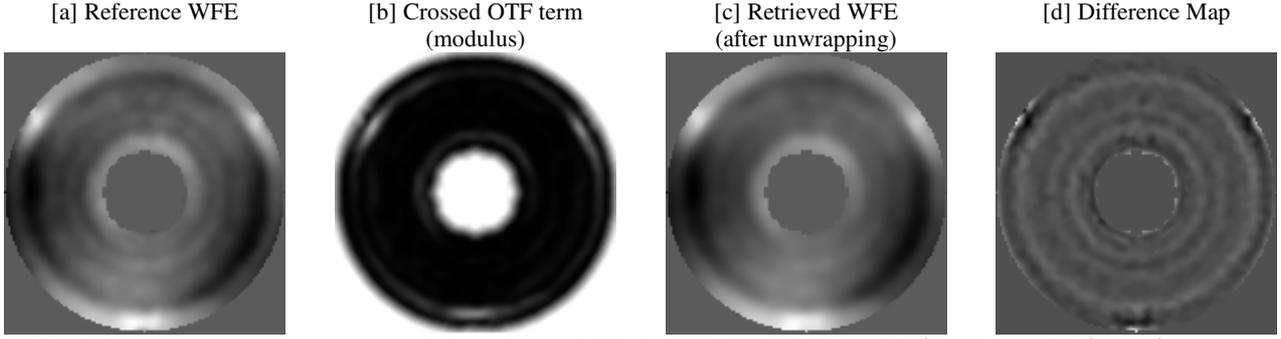

Fig.5. Case of high-spatial frequency polishing defects. [a] Reference WFE – PTV = 1.258 λ; RMS = 0.200 λ with λ = 0.6328 μm. [c] Reconstructed WFE – PTV = 1.182 λ; RMS = 0.195 λ. [d] Bi-dimensional difference-map – PTV = 0.347 λ; RMS = 0.018 λ. Grey-levels are scaled to PTV values.

## 2.2 Phase-shifting Telescope-Interferometer

Two major differences of the phase-shifting TI with respect to the off-axis version are indicated on the right side of Fig. 1: first, the reference pupil has been re-centred on the optical axis Z of the main telescope[1], and second, the whole reference surface can be moved along Z of known optical path quantities, corresponding to various phase shifts denoted $\phi$. The wave amplitude in the pupil plane now writes:

$$A(\vec{P}) = B_R(\vec{P})\exp[ik\Delta(\vec{P})] + B_r(\vec{P})\exp[i\phi] \tag{8}$$

Following the same reasoning than in section 2.1, and again employing the afore mentioned "Delta approximation" leads to a simplified expression of the OTF in the pupil plane, associated to a given phase-shift $\phi$:

$$C_\phi(\vec{P}) \approx OTF_R(\vec{P}) + C \times OTF_r(\vec{P}) + C \times B_R(-\vec{P})\exp[-ik\Delta(-\vec{P}) + i\phi] + C \times B_R(\vec{P})\exp[ik\Delta(\vec{P}) - i\phi] \tag{9}$$

Giving to $\phi$ successive values of 0, π/2, π and -π/2, a simple linear combination of the complex OTFs allows to retrieve the original phase, and therefore the Wavefront Error $\Delta(\vec{P})$.

$$B_R(\vec{P})\exp[ik\Delta(\vec{P})] \approx [C_0(\vec{P}) + iC_{\pi/2}(\vec{P}) - C_\pi(\vec{P}) - iC_{-\pi/2}(\vec{P})]/4C \tag{10}$$

Hence four different PSFs must be acquired here, whereas only one is necessary for the off-axis TI. Although this seems to be a severe drawback, advantages and limitations of both types of TIs are addressed in the next section. Numerical simulations presented in Fig. 6 illustrate the whole measurement sequence of the phase-shifting TI. The values of used parameters are similar to those of section 2.1, excepting Fried's radius $r_0$ that is here equal to 50 mm.

## 3. PERFORMANCE DISCUSSION AND LIMITATIONS

### 3.1 Error analysis

An extensive and in-depth analysis of all types of systematic errors or random noises that can affect TIs performance is beyond the scope of this communication. In view of building or prototyping such devices, this work will be necessary however, and was in fact already started in Ref. [15]. Table 1 provides a preliminary (and non exhaustive) list of errors, together with a summary of the main conclusions at this stage of the study. Two kinds of measurement uncertainties were distinguished, which are bias or systematic errors, on the one hand, and random noises, on the other hand. For what concerns bias errors, the major contributors proved to be the "Delta approximation" in Eqs. (7) and (9), and the useful spectral bandwidth and angular size (or magnitude) of the observed sky-object. Random errors, on their side, are fully dominated by the Signal-to-Noise Ratio (SNR) of the employed detector array, and more specially by photon noise. In particular, an analytical relationship between the SNR and subsequent WFE uncertainty $\delta\Delta(\vec{P})$ could be established:

$$|\delta\Delta(\vec{P})| < 1/(k \times SNR \times C) = S_R/(k \times SNR \times S_r) \tag{11}$$

---

[1] However this condition is not absolutely necessary: the reference pupil area could be de-centred with respect to the optical axis, provided that it is embedded in the main pupil area as in the optical configurations presented in section 4.

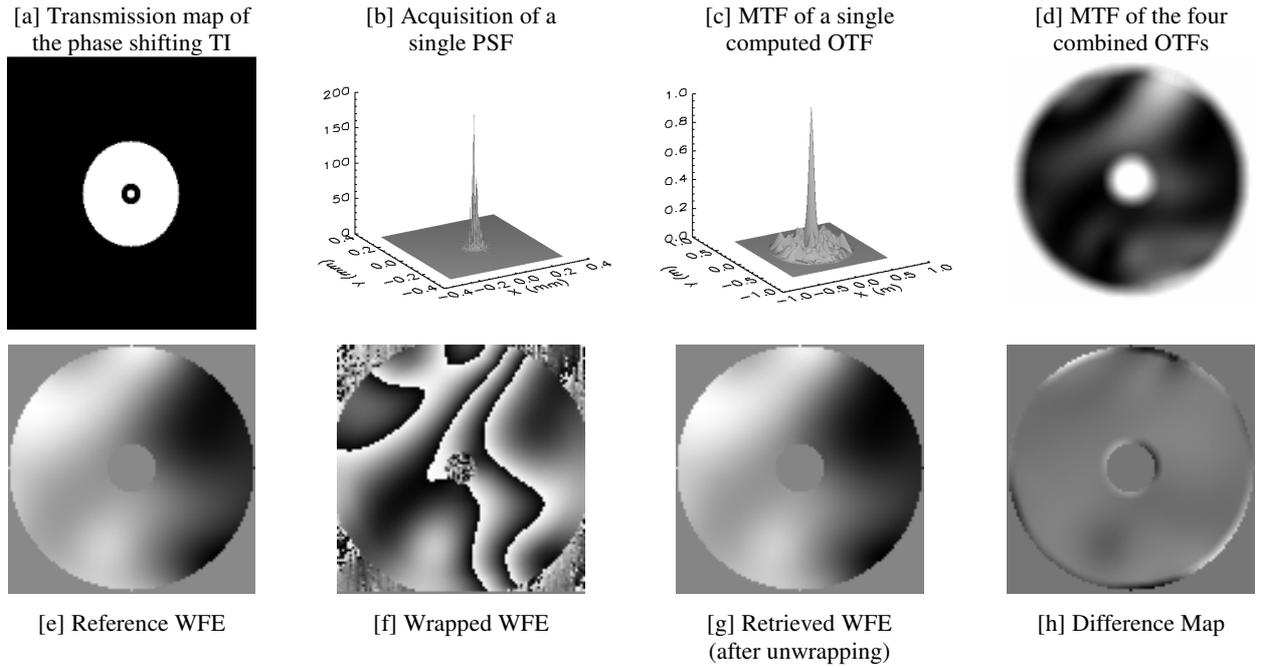

Fig. 6. Case of a phase shifting TI sensing atmospheric perturbations ($r_0$ = 50 mm). [a] Full transmission map of the TI, including the central reference pupil. [b] Acquisition of one single PSF. [c] MTF derived from one single acquired OTF. [d] MTF deduced from the four combined OTFs. [e] Reference WFE – PTV = 3.395 $\lambda$; RMS = 0.943 $\lambda$ with $\lambda$ = 0.6328 µm. [g] Reconstructed WFE – PTV = 3.281 $\lambda$; RMS = 0.937 $\lambda$. [h] Bi-dimensional difference-map – PTV = 0.251 $\lambda$; RMS = 0.017 $\lambda$. Grey-levels are scaled to PTV values.

Table 1. Non-exhaustive list of potential errors affecting Telescope-Interferometers and relevant lessons learned.

| TYPES OF ERRORS | LESSONS LEARNED |
|---|---|
| ***Systematic or Bias Errors*** | |
| Intrinsic error due to the "Delta approximation" | Can be maintained well below diffraction limit when $S_r \ll S_R$, but at the expense of increased photon noise. |
| Maximal spectral bandwidth | About 0.4 and 20 % for off-axis and phase-shift TIs respectively. |
| Angular radius of the observed sky-object | Corresponds to star magnitude ranging from –2 to 12, depending on TI type and dimensions. |
| Differential WFE between reference and main pupils | Negligible when reference pupil is diffraction-limited [12]. |
| Accuracy of phase steps (for phase-shift TIs) | Can be calibrated and corrected [16-17]. |
| ***Random Errors*** | |
| Detection noises including: | Mostly governed by photon noise – see Eq. (11) and section 4.2. |
| • Photon or shot noise | • Major contributor |
| • Read-out Noise (RON) | • Negligible |
| • Dark current | • Negligible |
| Atmospheric turbulence (for long integration times) | Method fails to retrieve WFEs. Should only be used on short integration times [15]. |
| Scintillation effect | Not studied so far. |

Relationship (11) represents indeed a fundamental formula for the dimensioning and performance assessment of any type of TI. Moreover, it shows that the attainable measurement accuracy $\delta\Delta(\vec{P})$ is linked to the spatial resolution $\Re$ of the device: the latter is indeed equal to the ratio $S_R/S_r$, which is the inverse of the TI contrast ratio C (see section 2). Hence Eq. (11) may be rewritten as:

$$|\delta\Delta(\vec{P})| < \Re/(k \times SNR) \qquad (12)$$

It can be concluded that for a given SNR of the detector, the higher is the desired spatial resolution, the worse the measurement errors will be. Conversely, low WFE spatial resolutions improve the intrinsic measurement accuracy of the TI system. Here a balance must be defined unambiguously between both the required spatial resolution and allowable measurement error: it can be felt intuitively that if the WFE to be estimated is mainly composed of low spatial frequency defects (e.g. pistons errors of a large segmented telescope, tip/tilt, or focus), larger values of r (and thus C) can be chosen, and the global accuracy $\delta\Delta(\vec{P})$ will be more favorable. An other important conclusion, however, is the fact that in order to perform seeing measurements in AO regime, the radius of the reference pupil should never exceed the actual Fried's radius $r_0$. Therefore two golden rules may be applied when dimensioning a TI:

- Select the highest $\Re$ satisfying condition $r < r_0$,
- Use Eqs. (11) or (12) in order to optimize the global measurement accuracy.

Running computer codes described in Ref. [15], TIs measurement errors were also estimated as a function of spectral bandwidth and angular radius of a target star. For that purpose, diameters of the main and reference pupils of both TIs (either off-axis or phase-shifting) were respectively set to D = 2R = 5 m and d = 2r = 1 m (the latter assumption implies excellent seeing conditions). Mean wavelength was $\lambda = 0.5$ μm and the primary mirror of the telescope was composed of seven or six hexagonal facets of 0.8 m side, each being affected with piston errors ranging from $-\lambda/2$ to $+\lambda/2$. SNRs were computed for an integration time $\tau = 10$ msec in order to stay compatible with an AO regime. Numerical results are plotted in Fig. 7. They show that

- Both designs radically differ for what concerns maximal spectral bandwidths. Off-axis TIs are found to be very sensitive to wavelength, since they cannot afford spectral ranges higher than 0.4 %. Conversely, phase shift TIs look much more convenient, since the maximal bandwidth is around 20 %.

- Maximal angular sizes of the observed star are around 20 mas, which correspond to magnitude –2. Photon noise dominates the off-axis TI whatever the star magnitude, while it governs phase shift TI above magnitude +1.

- Consequently, contribution of photon noise is so important that heavy numerical simulations such as those performed in this section become useless, and a realistic estimation of the TIs measurement accuracy can be obtained using the sole relationship (11). Thus all calculations of section 4.2 will be performed that way.

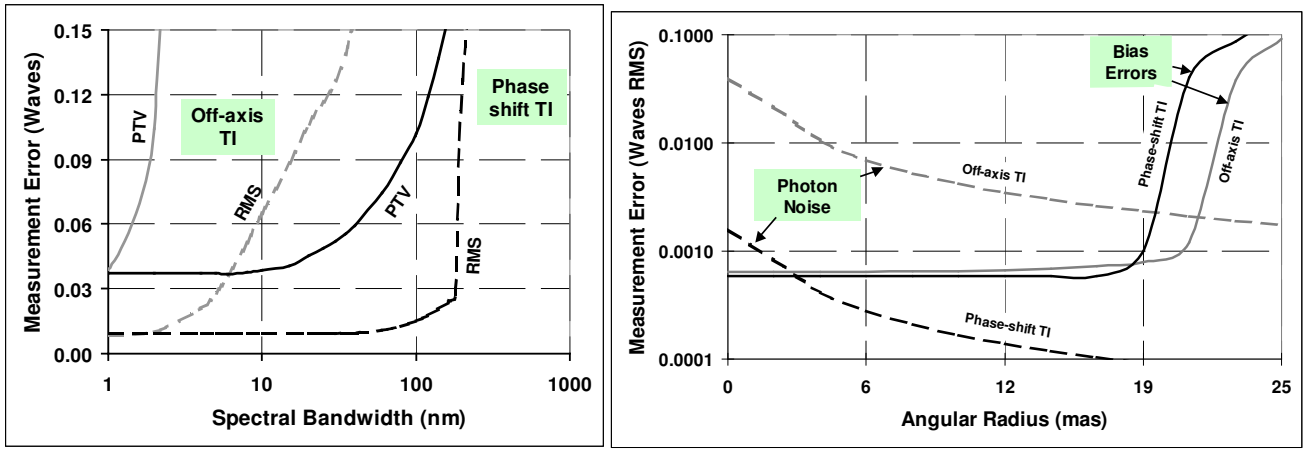

Fig. 7. Left, measurement accuracy as function of spectral bandwidth (black lines: phase-shifting TI; gray lines: off-axis TI; solid lines: PTV values; dashed lines: RMS values). Right, measurement accuracy as function of angular radius of the observed sky-object (solid lines: RMS bias error; dashed lines: photon noise contribution). The curves show undeniable superiority of phase-shifting TIs with respect to off-axis TIs.

## 3.2 Application range

Having asserted that photon noise is the dominant cause of error in a TI system, and knowing a simple relationship for evaluating it – Eq. (11) – allows to define more easily the application range of the method. The main scope of this section is to define acceptable limits on some critical parameters driving a TI performance, namely the main and auxiliary aperture diameters and the magnitude of the observed star. Here one of the main goals is to state if the method is appropriate for an AO system having the additional capacity of sensing the piston errors affecting the segmented mirrors of an ELT. Thus the main telescope diameter will be ranging from 10 to 50 meters, while reference pupil diameter varies from 0.1 to 1 meter. The wavelength $\lambda$ and integration time $\tau$ stay always equal to 0.5 µm and 10 msec respectively. We shall consider that photon noise is acceptable as long as it does not generate WFE uncertainties higher than the Maréchal's criterion, i.e. WFE < 0.075 $\lambda$ in RMS sense [14]. Some major conclusions were derived from the numerical results[1], which are illustrated in Fig. 8.

- The best measurement accuracies correspond to the smallest values of the main telescope diameter. This is indeed a direct consequence of Eq. (11), and the fact that the reference pupil area $S_r$ is limited by Fried's radius $r_0$. Therefore the best performance is attained for 10 m-class telescope diameter (left side of Fig. 8).

- The phase-shifting TI takes a clear advantage from its extended spectral range: a 10 m-diameter phase shift TI shows performance similar to a 50 m-diameter off-axis TI (left side of Fig. 8).

- An ELT of 30 m-diameter equipped with a phase-shift system should stay diffraction-limited when the reference pupil diameter $d = 2r_0$ is equal to 0.3 m. A 50 m-diameter ELT would require that $d = 2r_0 = 0.4$ m, which corresponds to good or very good seeing conditions (left side of Fig. 8).

- Depending on atmospheric disturbances, the limiting magnitude of a phase-shifting TI should vary between 8 ($r_0 = 0.25$ m) and 11 ($r_0 = 0.5$ m). The limiting magnitude of an off-axis TI could never exceed 4 (right side of Fig. 8).

Hence the phase-shifting TI looks the most promising. However it still suffers from an incomplete sky coverage, on the one hand, and only shows its best performance when atmospheric seeing is favorable, on the other hand. For those two reasons it cannot be considered as part of a multi-purpose AO system, but could be advantageously employed in some particular circumstances (periodical co-phasing of reflective facets on an ELT, scientific observations of moderate magnitude sky-objects, such as those achieved by "planet-finding" instruments).

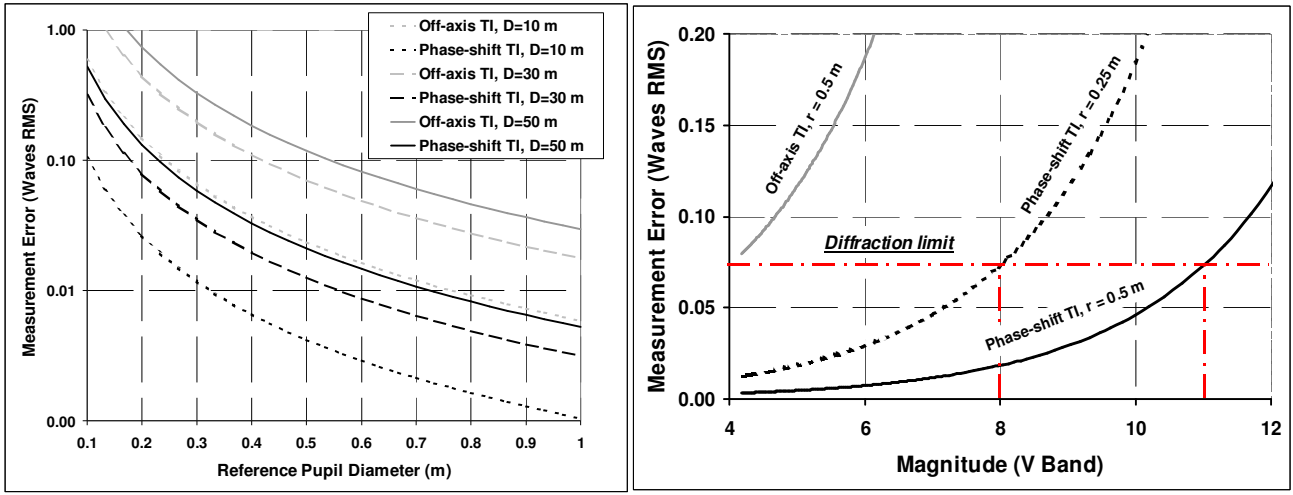

Fig. 8. Left, measurement accuracy as function of reference pupil diameter $d = 2r$, for various telescope diameters $D = 2R$ (black lines: phase-shifting TI; gray lines: off-axis TI; dotted lines: $D = 10$ m; dashed lines: $D = 30$ m; solid lines: $D = 50$m). Right, measurement accuracy as function of star magnitude for $D = 30$ m (dashed line: $r = 0.25$ m; solid lines: $r = 0.5$ m). Here again advantages of the phase-shifting TI clearly appear.

---

[1] Although some results do not fully comply with those presented in Ref. [15], major conclusions remain unchanged.

### 3.3 Telescope-Interferometers trade-off

The following criteria were selected in order to establish a preliminary trade-off between both types of TIs: spectral bandwidth, limiting magnitude of the target star, minimal number of required pixels[1], and hardware costs and complexity. Elements of answer are summarized in Table 2.

Table 2. Telescope-Interferometers trade-off.

| Criteria | OFF-AXIS TI | PHASE-SHIFTING TI |
|---|---|---|
| Allowed spectral bandwidth | $\leq 0.4\%$ | Between 20 and 30 % |
| Limiting star magnitude for D = 30 m | 4 | 11 |
| Minimal number of pixels | $1024 \times 1024$ | $256 \times 256$ |
| Hardware complexity and costs | High, requires manufacturing of dedicated auxiliary telescope and associated delay line [12] | Moderate, can be implemented with small and simple optical components located at the telescope focal plane (see section 4) |

The conclusion of this trade-off seems obvious, since the phase-shifting TI is superior from any point of view. This leads to finally discard the off-axis TI from further design study – even preliminary. Therefore the following section 4 will only be focused at the practical implementation of a phase-shifting Telescope-Interferometer.

## 4. PRACTICAL IMPLEMENTATION ON A TELESCOPE FACILITY

In this section are described two possible optical arrangements for wavefront sensing based on the principle of phase-shifting TIs. Those measurement schemes are new, since previously proposed implementations in Refs. [12] and [13] were well founded, but suffered from a few practical drawbacks[2]. WFE measurements can indeed be performed following two different modes, depending on PSFs acquisition schemes. The latter can either be simultaneous or sequential, as described in sections 4.1 and 4.2 respectively. In both cases the phase sensing device is located behind the telescope focal plane, within a compact optical layout where the four phase-shifts $\phi = 0$, $\pi/2$, $\pi$ and $-\pi/2$ are added to the telescope WFE. It is also assumed that the primary mirror of the telescope includes a "reference segment" of high image quality (i.e. diffraction-limited), corresponding to the reference pupil area where phase-shifts have to be introduced (see Figs. 9 and 10).

### 4.1 Simultaneous measurements

Here PSFs acquisitions are realized simultaneously, during a full integration time $\tau = 10$ msec. The WFE sensing device incorporates one Collimating Lens (CL1, see Fig. 9), then splits the collimated beam into four different optical arms by means of three beam-splitters BS1, BS2 and BS3. Each optical arm is composed of the following optics or electronics components:

- A Phase Plate (denoted PP1, PP2, PP3 and PP4 in Fig. 9) in charge of adding the reference phase-shift $\phi$ to the telescope WFE. The Phase Plate is located at an image plane of the telescope pupil, near CL1 focal plane. The phase-shift $\phi$ is introduced by means of an equivalent glass thickness in the reference pupil area. Depending on the selected material and effective spectral bandwidth of the device, it might be necessary – or not – to use a set of achromatic plates similar to those currently used in nulling interferometry [18], instead of one single plate.

- A Focusing Lens (FL1, FL2, FL3 and FL4) re-imaging the phase-shifted PSF on a detector array. The focal length of the FL is adjusted in order to achieve a certain magnification ratio M between telescope and camera focal planes. The exact value of M depends on the camera pixels size and the required OTF spatial sampling.

- A CCD detector array (Cameras 1-4 in Fig. 9) finally acquiring the phase-shifted PSF.

---

[1] Since a limited pixels number is more suited to implementation of the algorithms (e.g. FFT) in quasi real-time.
[2] Let us mention for example the "Michelson configuration" in Ref. [12], or the "alternative design" of Ref. [13], section 4.2, where the reference pupil was integrated into the segmented primary mirror of the main telescope: the major difficulty was here to control the displacement of the large reference mirror with sufficient precision.

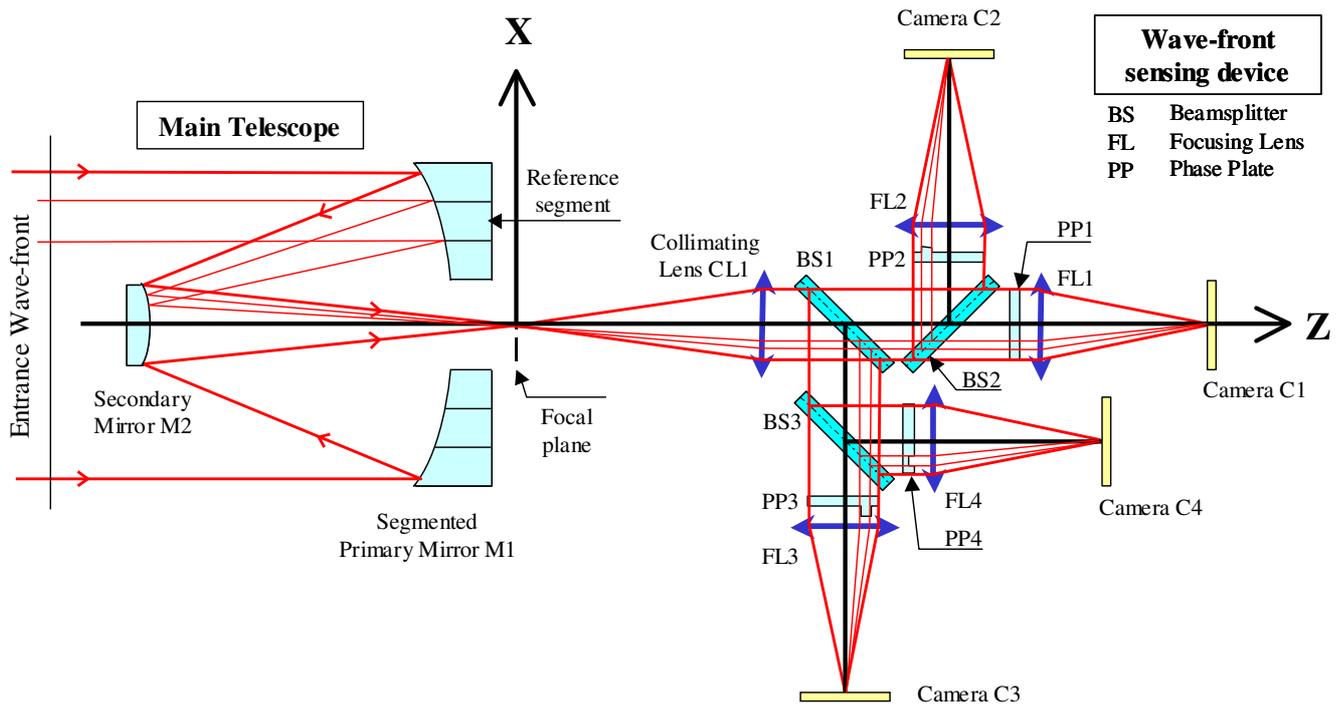

Fig. 9. Schematic view of a phase-shifting TI arrangement, designed for simultaneous PSF/OTF measurements.

The major advantage of this WFS configuration consists in the simultaneity of PSFs measurements, ensuring a reliable WFE reconstruction. In return, the presence of several beam-splitters will decrease the signal levels recorded by each camera of at least 75%, which represents a strong disadvantage for astronomical applications. The following configuration eliminates this drawback, since PSF measurements are realized sequentially.

### 4.2 Sequential measurements

In this configuration, one single CCD camera is needed for the four different PSF acquisitions. Measurements are performed sequentially rather than simultaneously. The full acquisition time must stay equal to 10 msec, thus the elementary integration time (at each different phase-shift $\phi$) will be $\tau = 2.5$ msec. As in previous section, the WFE sensing device is composed of one Collimating Lens imaging the telescope exit pupil on a reference flat mirror (see Fig. 10). That mirror is pierced at a location corresponding to the telescope reference facet, and a mandrel carrying a small, flat optical surface is piezoelectrically moved along the optical axis, thus generating the required phase-shifts $\phi$. A Focusing Lens finally forms consecutive images of the PSF in the plane of the detector array, where they are recorded before data processing. From a radiometric point of view, this last configuration appears as the most favorable, since resulting SNRs should be multiplied by a factor around two. On the other hand, any change of the telescope WFE between successive acquisitions may alter the reconstruction process and then the final achieved accuracy.

### 5. CONCLUSION

This paper provides a synthesis about the principles, performance and limitations of what I named "Telescope-Interferometers" in previous works – Refs. [12-13] and [15]. The basic idea consists in transforming a telescope into a WFE sensing device by giving to sky photons an additional access to the focal plane of the telescope through its exit pupil. This can be achieved in two different ways, namely the off axis and phase-shifting TIs. In the first case a small and decentred reference telescope is added aside the main pupil, creating a weakly modulated fringe pattern in the image plane. In the second configuration, different calibrated phase shifts are applied over the reference pupil area. In both cases the PSFs measured in the focal plane of the telescope carry information about the transmitted WFE, which is retrieved via fast and simple algorithms suitable to an adaptive optics operational mode.

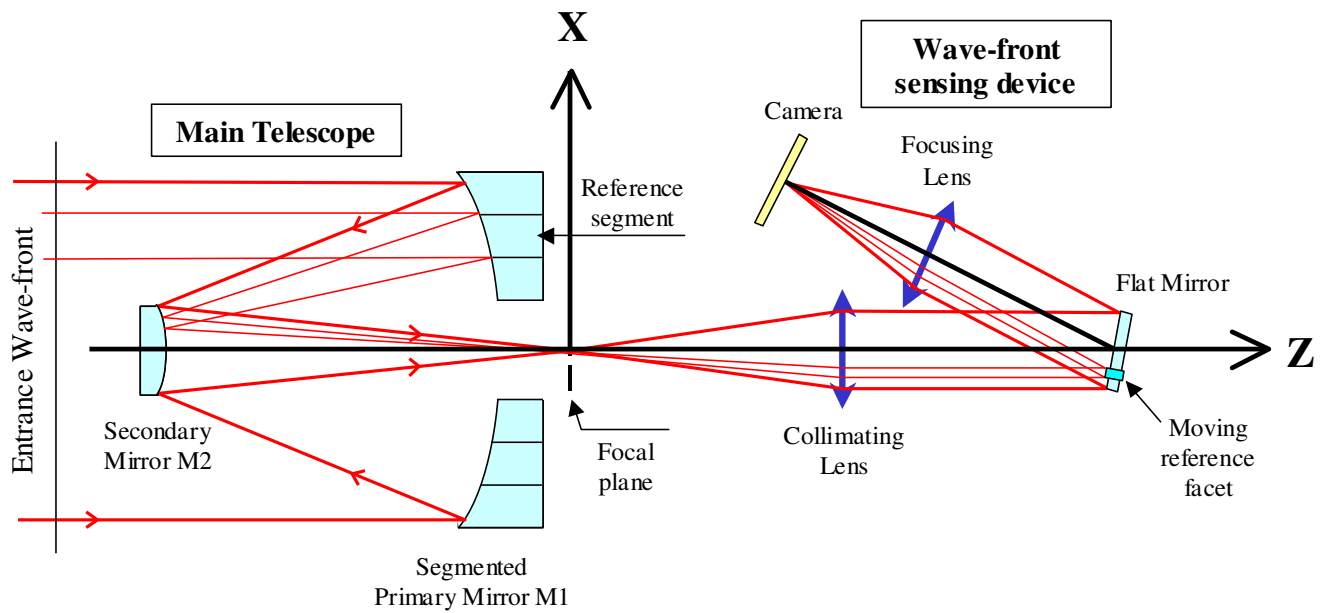

Fig. 10. Schematic view of the phase-shifting TI arrangement, designed for sequential PSF/OTF measurements.

Herein was evaluated the accuracy of both types of TIs, in terms of noise and systematic errors. It was highlighted that RMS measurement error is proportional to the geometrical area of the main telescope, and inversely proportional to detector SNR and reference pupil area. The spectral bandwidth $\Delta\lambda$ and radiating properties of the target star were defined (e.g. $\Delta\lambda < 150$ nm and magnitude comprised between -2 and +11 for a phase-shifting TI working in the V band). It was shown that WFE measurement errors are particularly sensitive to photon noise, which rapidly governs the achieved accuracy for telescope diameters higher than 10 m. Nevertheless, TI method seems to be applicable to adaptive optics systems on telescope diameters ranging from 10 to 50 m (i.e. ELTs), depending on the seeing conditions and magnitude of the observed stars. Also, phase-shifting TIs were found clearly superior to their off-axis version – assuming the same geometrical characteristics. This led to rule out the off-axis TI method at the end of a short trade-off, and only tentative designs based on phase-shifting TI principle were presented in the last section, showing two promising wavefront sensing configurations.

Other attractive theoretical studies remain to be undertaken around TI matter, such as application to scintillation measurements, comparison with different types of WFS – and particularly those depicted in Refs. [10-11] – or improvements of the method by implementing a deconvolution process in order to get rid of the "Delta approximation". However, priority should be given now to the practical realization of an experiment or a TI prototype, whose results would validate the theory, predicted performance and errors estimation. This seems to be a mandatory step before a larger-scale implementation of a phase-shifting TI can be envisaged on an existing telescope facility.

# REFERENCES

[1] H.W. Babcock, "The Possibility of Compensating Astronomical Seeing," Pub. Astron. Soc. Pacific vol. 65, p. 229-236 (1953).

[2] T. Fusco, G. Rousset, J.F. Sauvage, C. Petit, J.L. Beuzit, K. Dohlen, D. Mouillet, J. Charton, M. Nicolle, M. Kasper, P. Baudoz and P. Puget, "High-order adaptive optics requirements for direct detection of extrasolar planets: Application to the SPHERE instrument," Optics Express, vol. 14, p. 7515-7534 (2006).

[3] A. Labeyrie, "Resolved imaging of extra-solar planets with future 10-100 km optical interferometric arrays," AAS Ser. Vol. 118, p. 517-524 (1996).

[4] R.G. Lane and M. Tallon, "Wave-front reconstruction using a Shack-Hartmann sensor," Appl. Opt. vol. 31, p. 6902-6908 (1992).

[5] F. Roddier, "Curvature sensing and compensation: a new concept in adaptive optics," Appl. Opt. vol. 27, p. 1223-1225 (1988).

[6] R. Ragazzoni, "Pupil plane wavefront sensing with an oscillating prism," Journal of Modern Optics vol. 43, p. 289-293 (1996).

[7] F. Hénault, "Wavefront sensor based on varying transmission filters: theory and expected performance," Journal of Modern Optics vol. 52, p. 1917-1931 (2005).

[8] J.R. Fienup, "Phase retrieval algorithms: a comparison," Appl. Opt. vol. 21, p.2758-2769 (1982).

[9] R.A. Gonsalves, "Phase retrieval and diversity in adaptive optics," Optical Engineering vol. 21, p. 829-832 (1982).

[10] R. Angel, "Imaging extrasolar planets from the ground," in Scientific Frontiers in Research on Extrasolar Planets, D. Deming and S. Seager eds., ASP Conference Series vol. 294, p. 543-556 (2003).

[11] A. Labeyrie, "Removal of coronagraphy residues with an adaptive hologram, for imaging exo-Earths," in Astronomy with High Contrast Imaging II, C. Aime and R. Soummer eds., EAS Publications Series vol. 12, p. 3-10 (2004).

[12] F. Hénault, "Analysis of stellar interferometers as wavefront sensors," Appl. Opt. vol. 44, p. 4733-4744 (2005).

[13] F. Hénault, "Conceptual design of a phase shifting telescope-interferometer," Optics Communications vol. 261, p. 34-42 (2006).

[14] A. Maréchal and M. Françon, "Diffraction, structure des images", Masson & Cie eds., 120, Boulevard Saint-Germain, Paris VIe (1970).

[15] F. Hénault, "Signal-to-noise ratio of phase sensing telescope interferometers," J. Opt. Soc. Am. A vol. 25, p. 631-642 (2008).

[16] J. Schwider, R. Burow, K.E. Elssner, J. Grzanna, R. Spolaczyk and K. Merkel, "Digital wave-front measuring interferometry: some systematic error sources," Appl. Opt. vol. 22, p. 3421-3432 (1983).

[17] K. Kinnstaetter, A.W. Lohmann, J. Schwider and N. Streibl, "Accuracy of phase shifting interferometry," Appl. Opt. vol. 27, p. 5082-5089 (1988).

[18] Y. Rabbia, J. Gay, J.P. Rivet and J.L. Schneider, "Review of Concepts and Constraints for Achromatic Phase Shifters," in Proceedings of GENIE-DARWIN Workshop - Hunting for Planets, H. Lacoste ed., ESA SP-522 (European Space Agency, 2003), 7.1.